\documentclass[aps, reprint, prb, amsfonts, amssymb, preprintnumbers, showpacs, notitlepage, superscriptaddress, floatfix]{revtex4-1}

\usepackage[tbtags]{amsmath}
\usepackage{graphicx}
\usepackage{bm}
\usepackage{dcolumn}
\usepackage{epsfig}
\usepackage[caption = false]{subfig}
\usepackage{multirow}
\usepackage{textcomp}
\usepackage{mathtools}

\usepackage{comment}

\newcommand{\vQ}{{\mbox{\boldmath$Q$}}}

\newcommand{\rhoqx}{{\mbox{$\rho_{2{\bf Q}_x}$}}}
\newcommand{\rhoqy}{{\mbox{$\rho_{2{\bf Q}_y}$}}}

\newcommand{\dq}[1][{}]{\ensuremath{\Delta_{\bm{Q}_{#1}}}}
\newcommand{\ndq}[1][{}]{\ensuremath{\Delta_{-\bm{Q}_{#1}}}}
\newcommand{\Dxdq}[1][{}]{\ensuremath{D_x}\dq[#1]}
\newcommand{\Dydq}[1][{}]{\ensuremath{D_y}\dq[#1]}

\newcommand{\Cydq}[1][{}]{\ensuremath{(\Dydq[#1])^*}}
\newcommand{\Dzdq}[1][{}]{\ensuremath{D_z}\dq[#1]}
\begin{document}

\title{Emergent loop current order from pair density wave superconductivity}
\author{D. F. Agterberg}
\affiliation{Department of Physics, University of Wisconsin-Milwaukee, Milwaukee, WI 53211, USA}
\author{Drew S. Melchert}
\affiliation{Department of Physics, University of Wisconsin-Milwaukee, Milwaukee, WI 53211, USA}
\affiliation{Department of Physics, University of Chicago, Chicago, IL 60637, USA}
\author{M. K. Kashyap}
\affiliation{Department of Physics, University of Wisconsin-Milwaukee, Milwaukee, WI 53211, USA}
\date{\today}
\begin{abstract}
There is evidence that the pseudogap phase in the cuprates breaks time-reversal symmetry. Here we show that pair density
wave (PDW) states give rise to a translational invariant nonsuperconducting order parameter that breaks time-reversal and parity symmetries, but preserves their 
product. This secondary order parameter has a different origin, but shares the same symmetry properties as a magnetoelectric loop current order that has been proposed 
earlier in the context of the cuprates to explain the appearance of intracell magnetic order. We further show that, due to fluctuations, this secondary loop current 
order, which breaks only discrete symmetries, can preempt PDW order, which breaks both continuous and discrete symmetries. In such a phase, the emergent loop current
order coexists with spatial short-range superconducting order and possibly short-range charge density wave (CDW) order. Finally, we propose a PDW phase that accounts 
for intracell magnetic order and the Kerr effect, has CDW order consistent with x-ray scattering and nuclear magnetic resonance observations, and quasi-particle 
properties consistent with angle-resolved photoemission scattering.
\end{abstract}
\maketitle
\section{\label{sec: intro} Introduction}
A central question in the underdoped cuprate superconductors is the origin of the pseudogap phase. This phase was originally thought to to be a precursor phase to
superconductivity with spin-singlet pairs, no phase coherence, and no broken symmetries \cite{eme95,lee06}.  However, more recent measurements suggest broken symmetries.
Specifically, polarized elastic neutron scattering observe intraunit cell magnetic order \cite{sid13} at a temperature close to the onset of a polar Kerr
effect \cite{xia08,he11} (see also Ref.~\onlinecite{kam02}). This suggests broken time-reversal symmetry \cite{var97,cha01}. Also, static quasi-long-range charge 
density wave (CDW) order has been observed through x-ray scattering \cite{ghi12,com13,sil13} and through nuclear magnetic resonance \cite{wu14}. This order appears at 
the incommensurate wavevectors $2{\bm Q}_x=(2Q,0)$ and $2{\bm Q}_y=(0,2Q)$ \cite{com13}. In addition, there exists evidence for superconducting (SC) correlations in the 
pseudogap phase. Diamagnetism is observed much above $T_c$ \cite{li10} and also at fields that far exceed the estimated mean-field SC upper critical field \cite{yu14}. 
To explain the prevalence of SC correlations and CDW order, pair density wave (PDW) order has been suggested as an order parameter for the pseudogap phase 
\cite{lee14,yu14}. This proposal was bolstered by a demonstration that PDW order accounts for anomalous quasi-particle (qp) properties observed by angle-resolved 
photoemission (ARPES) \cite{lee14}. PDW superconductivity is a spatially varying SC state similar to  Fulde Ferrell Larkin Ovchinnikov (FFLO) states \cite{ful64,lar65}. 
It has been discussed in a variety of contexts for the cuprates \cite{ber09,agt08,cor14,zel11,lee14}.
%

\begin{figure}[t]
\begin{center}
\includegraphics[width=2.55in]{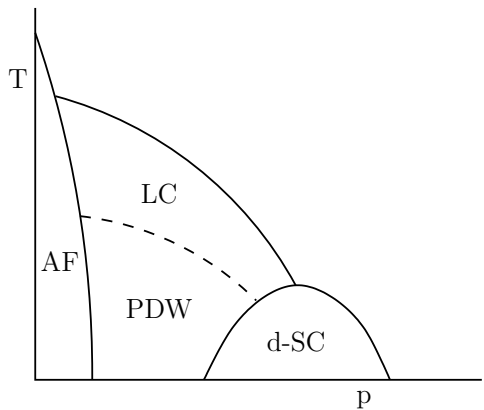}
\end{center}
\caption{Qualitative temperature ($T$) versus hole doping ($p$) phase diagram. Here LC represents the ME loop current phase, PDW represents the pair density wave phase, 
AF represents antiferromagnetism, and d-SC represents $d$-wave superconductivity.} 
\label{fig1}
\vspace{-4mm}
\end{figure}

Here we show that PDW order can naturally induce a translational invariant secondary order parameter that breaks both time-reversal and parity symmetries, but is
invariant under the product of the two. Similar order parameters with this symmetry have appeared in the context of the cuprates under the name magnetoelectric (ME)
order \cite{ore11} and as ME loop current order \cite{sim02}. Here we name such order ME loop current order. We further show that there exists a mean-field PDW ground
state with ME loop current order that accounts for the Kerr effect and for intracell magnetic order, with CDW order at the observed wavevectors, and which accounts for
qp properties observed by ARPES \cite{he11}. This PDW ground state has continuous  $U(1)$ degeneracies (associated with broken SC gauge and translational symmetries)
together with a discrete degeneracy associated with the ME loop current order. Fluctuations of the $U(1)$ degeneracies suppress both the SC and CDW order, allowing for
a state with spatial long-range ME loop current order and short-range SC and CDW orders (Fig.~\ref{fig1}). We propose that this state is responsible for behavior that 
emerges at the pseudogap temperature $T^*$ \cite{he11}. Such a ME loop current state is conceptually similar to the nematic phase that arises due to magnetic 
fluctuations proposed for the pnictides \cite{fer12} and to a translational invariant broken time-reversal symmetry state stemming from CDW and modulated bond current 
orders \cite{wan14}.

Since it is closely related to ME loop current PDW state we find, and has been used to explain the anomalous qp properties observed through ARPES experiments, we
highlight the recent PDW proposal of Lee \cite{lee14}. In particular, this proposal has its origin in a gauge theory description of the resonating valence bond phase.
Here, pairing occurs through a transverse gauge field and leads to an incommensurate checkerboard PDW state for which the PDW order can be qualitatively expressed as
$\Delta({\bm x})=\Delta_Q[\cos({\bm Q}_x\cdot {\bm x})+i\cos({\bm Q}_y\cdot {\bm x})]$. This state has secondary CDW order at wavevectors $2{\bm Q}_x$ and $2{\bm Q}_y$,
in agreement with experiment. This state cannot account for the observed signatures of translational invariant broken time-reversal symmetry.

In the following, we begin with a summary of the symmetry properties of PDW order and introduce the translational invariant loop current order parameter. This is followed 
by the relevant PDW action for tetragonal symmetry. For tetragonal symmetry, it is not possible to analytically find all possible ground states. For this reason we then 
turn to an analysis of PDW order for a theory with orthorhombic symmetry. This theory allows for a complete understanding of all allowed PDW ground states and can be used 
to establish the existence of a phase which has long-range translation invariant loop current order but no long-range superconducting or CDW order.  We then return to 
tetragonal symmetry and examine a loop current phase that is a natural generalization of that found for orthorhombic symmetry. After this we show there exists a PDW state 
that shares the same symmetry properties as the recent tilted loop current phase discussed by Yakovenko \cite{yak14}. This phase is consistent with all observations of 
broken time-reversal symmetry in the underdoped cuprates.  Finally, we examine the quasi-particle (qp) properties relevant to ARPES measurements for the tetragonal ME 
PDW phase. We show that while the qp properties of the ME PDW phase are similar to those found by Lee \cite{lee14} for a PDW phase without loop current order, there are 
observable differences that will allow these two phases to be distinguished.
\vspace{-6mm}
\section{\label{sec: PDW-loop} PDW induced translational invariant loop current order}
\vspace{-2mm}
PDW order originates when paired fermions have a finite center of mass momentum. It is characterized by order parameter components $\dq$ which, under a translation
$\bm{T}$, transform as  $\dq\rightarrow e^{i\bm{T}\cdot \bm{Q}}\dq$. Key here are the transformation properties under time-reversal $\mathcal{T}$
and parity symmetries $\mathcal{P}$:

\begin{equation}
\dq\xrightarrow{\mathcal{T}} \Delta^*_{-\bm Q} \qquad \qquad \dq\xrightarrow{\mathcal{P}} \ndq.
\end{equation}
These symmetries suggest a consideration of the secondary ME loop current order parameter $l= (|\dq[i]|^2-|\ndq[i]|^2)$. This order parameter has translational
invariance, is odd under both  $\mathcal{T}$ and $\mathcal{P}$, and invariant under the product $\mathcal{T}\mathcal{P}$. If a PDW ground state satisfies
$|\dq[i]| \ne |\ndq[i]|$, then the state will have non-zero $l$. This condition is not satisfied by any of the PDW states proposed in the context of the cuprates
\cite{ber09,cor14,lee14,yu14}. This motivates the question, are there stable PDW ground states that do exhibit loop current order? Below we show there are.
We find that there exists a PDW ground state that can qualify as a pseudogap mean-field order parameter. We impose the following four criteria on such a state:
\vspace{0.1cm}

\noindent 1- It is a mean-field ground state of a Ginzburg-Landau-Wilson (GLW) action (for parameters that are not a set of measure zero in the GLW action parameter 
space).\vglue 0.2 cm
\noindent 2- It has finite $l$ and accounts for the Kerr effect and intracell magnetic order.\vglue 0.2 cm
\noindent 3- It has CDW correlations at the observed momenta.\vglue 0.2 cm
\noindent 4- It can account for ARPES spectra. \vspace{0.1cm}

Prior to defining the PDW order parameter we consider in more detail, it is useful to point out that there are two previously found PDW ground states that should have
finite $l$. The first is the well known Fulde-Ferrel (FF) phase for which $\Delta({\bm x})=e^{i{\bm Q}\cdot {\bm x}}$. This state has no CDW order and therefore cannot
represent a pseudogap order parameter. The second state is found in Ref.~\onlinecite{agt08}, for which the gap can qualitatively be represented as
$\Delta({\bm x})=\dq{[e^{i{\bm Q}_x\cdot {\bm x}}+e^{i{\bm Q}_y\cdot {\bm x}}]}$.  This state has CDW order, but this order is not at a wavevector that matches 
experiment and, consequently, cannot be a pseudogap order parameter.

Criterion 4 strongly restricts our search for a pseudogap order parameter. Specifically, we require that the Fermi arc is reproduced, the low energy bands near the
anti-nodal point are reproduced (which has a gap minimum at momentum $k_G\ne k_F$, where $k_F$ is Fermi momentum) \cite{he11}, and the Fermi arc is derived
from occupied states moving up towards the Fermi energy \cite{he11,lee14}. The PDW state discussed in Ref.~\onlinecite{lee14} gives rise to these properties, and
it is natural to use this as a starting point. However, the GLW theory based on the PDW momenta chosen in Ref.~\onlinecite{lee14} does not produce a ground state that
satisfies the above four criteria and we must therefore consider generalizations of this state. To identify such a generalization, we note that 
a key feature of Ref.~\onlinecite{lee14} that allows the ARPES spectra to be
reproduced is the choice of the momenta about which fermions are paired. In particular, the mean-field pairing Hamiltonian for PDW order is
\begin{equation} 
H=\sum_{{\bm p},s} \epsilon_{{\bm p}}c^{\dagger}_{{\bm p}s} c_{{\bm p}s}+ \sum_{{\bm Q}_i,{\bm p}}[\Delta_{{\bm Q}_i}({\bm p})c^{\dagger}_{{\bm p}+{\bm K}_i\uparrow} 
c^{\dagger}_{-{\bm p}+{\bm K}_i,\downarrow}\\+h.c.],
\end{equation}
where $c_{{\bm k}s}$ is the fermion destruction operator with momentum ${\bm k}$ and spin $s$, $\epsilon_{\bm k}$ is the bare dispersion, and  $h.c.$ means Hermitian
conjugate. The momenta about which the fermions are paired are the ${\bm K}_i$, leading to PDW order at ${\bm Q}_i=2{\bm K}_i$. In the following we examine PDW order
that stems from the ${\bm K}_i$ shown in Fig.~2.
\begin{figure}[t]
\begin{center}
\includegraphics[width=2.55in]{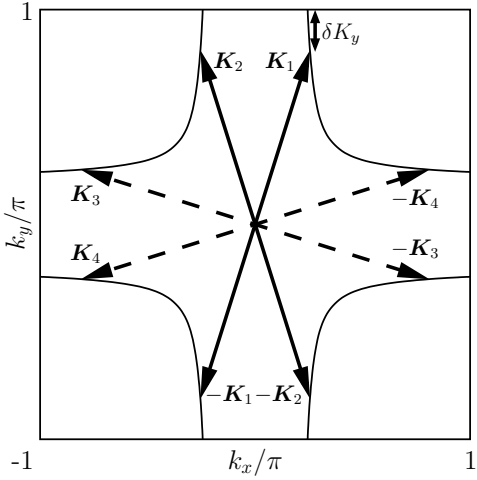}
\end{center}
\caption{\label{fig: PDW-momenta}The positions of the momenta $\bm {K}_i$ about which PDW Cooper pairs are formed. The corresponding eight PDW order parameter 
components $\dq[i]$ have momenta $\bm{Q}_i=2\bm{K}_i$. The solid line momenta apply only to the theory with orthorhombic symmetry, and all the momenta (solid and 
dashed) are included for tetragonal symmetry. The displacement $\delta K_y$ denotes the shift of the momenta $\bm{K}_i$ from the zone edge. When $\delta K_y=0$, the 
theory of Ref.~\onlinecite{lee14} is reproduced.}
\vspace{-4mm}
\end{figure}
In the limit that $\delta K_y=0$, the theory of Ref.~\onlinecite{lee14} is reproduced. Consequently, for sufficiently
small $\delta K_y$, the PDW states examined here should be able to reproduce the ARPES spectra. We show that this is indeed the case in Section \ref{sec: quasi}.
\vspace{-4mm}
\section{\label{sec: GLW-tetra} GLW Action: tetragonal symmetry}
The momenta specified in Fig.~\ref{fig: PDW-momenta} lead to a PDW order parameter with eight complex degrees of freedom:  $(\dq[1], \dq[2], \dq[3],\dq[4],\ndq[1], 
\ndq[2], \ndq[3], \ndq[4])$. To construct the GLW free energy, the transformation properties of this order parameter under rotations are required. The point group 
symmetry is $D_{4h}$ with generators $\{ C_4, \sigma_{x}, \sigma_z \}$  where $C_4$ is a 4-fold rotation about the $c$-axis and $\sigma_x$ ($\sigma_z$) is a mirror
reflection through $y$-$z$ ($x$-$y$) plane. Under these generators,  the PDW order $(\dq[1], \dq[2], \dq[3],\dq[4],\ndq[1], \ndq[2], 
\ndq[3], \ndq[4])$ transforms as
\begin{equation}
\begin{aligned}
C_4:& (\dq[3], \dq[4], \ndq[1], \ndq[2], \ndq[3], \ndq[4], \dq[1], \dq[2]),\\
\sigma_{x}:& (\dq[2], \dq[1], \ndq[4], \ndq[3], \ndq[2], \ndq[1], \dq[4], \dq[1]),\\
\sigma_z:& (\dq[1], \dq[2], \dq[3],\dq[4], \ndq[1], \ndq[2], \ndq[3], \ndq[4]).
\end{aligned}
\end{equation}
Considering invariance under translations, rotations, time-reversal, parity and gauge symmetries, the
corresponding GLW action can be written as: $S_{0, \text{tet}} = S_{0, \text{hom}} + S_{0, \text{grad}}$. Here, $S_{0, \text{hom}}$ and $S_{0, \text{grad}}$ are
\begin{widetext}
\begin{multline}\label{eq: tetra-free1_hom}
  S_{0,\text{hom}} = r_0 \sum\nolimits_i |\dq[i]|^2 + \beta_1 {\left( \sum\nolimits_i |\dq[i]|^2 \right)}^2 \\
     + \beta_2 \left( |\dq[1]|^2 |\ndq[1]|^2 + |\dq[2]|^2 |\ndq[2]|^2 + |\dq[3]|^2 |\ndq[3]|^2 + |\dq[4]|^2 |\ndq[4]|^2\right)\\
     + \beta_3 \left( |\dq[1]|^2 |\dq[2]|^2 + |\dq[3]|^2 |\dq[4]|^2 + |\ndq[1]|^2 |\ndq[2]|^2 + |\ndq[3]|^2 |\ndq[4]|^2 \right) \\
     + \beta_4 \left( |\dq[1]|^2 |\dq[3]|^2 + |\dq[2]|^2 |\dq[4]|^2 + |\dq[3]|^2 |\ndq[1]|^2 + |\dq[4]|^2 |\ndq[2]|^2 \phantom{~~~~.} \right.\\
	\shoveright{ \left. + \phantom{.}|\ndq[1]|^2 |\ndq[3]|^2 + |\ndq[2]|^2 |\ndq[4]|^2 + |\ndq[3]|^2 |\dq[1]|^2 + |\ndq[4]|^2 |\dq[2]|^2 \right)} \\
     + \beta_5 \left( |\dq[1]|^2 |\dq[4]|^2 + |\ndq[1]|^2 |\ndq[4]|^2 + |\dq[2]|^2 |\ndq[3]|^2 + |\dq[3]|^2 |\ndq[2]|^2  \right)\\
     + \beta_6 \left( |\dq[2]|^2 |\dq[3]|^2 + |\dq[4]|^2 |\ndq[1]|^2 + |\ndq[2]|^2 |\ndq[3]|^2 + |\ndq[4]|^2 |\dq[1]|^2 \right) \\
     + \beta_7 \left( |\dq[1]|^2 |\ndq[2]|^2 + |\dq[2]|^2 |\ndq[1]|^2 + |\dq[3]|^2 |\ndq[4]|^2 + |\dq[4]|^2 |\ndq[3]|^2 \right) \\
     + \beta_{c_1} \left\{ \left[\dq[1]\ndq[1](\dq[2]\ndq[2])^*  + \dq[3]\ndq[3](\dq[4]\ndq[4])^* \right ] + c.c.  \right\}	\phantom{~~~~~~~~~~~..}	\\
     + \beta_{c_2} \left\{ \left[\dq[1]\ndq[1](\dq[3]\ndq[3])^*  + \dq[2]\ndq[2](\dq[4]\ndq[4])^* \right ] + c.c.  \right\}	\phantom{~~~~~~~~~~~..}	\\
     + \beta_{c_3} \left\{ \left[\dq[1]\ndq[1](\dq[4]\ndq[4])^*  + \dq[2]\ndq[2](\dq[3]\ndq[3])^* \right ] + c.c.  \right\},	\phantom{~~~~~~~~~~~~~~~~~~~~~~~~~~~.....}
\end{multline}
\begin{multline}\label{eq: tetra-free1_grad}
S_{0, \text{grad}} = \kappa_1 \sum\nolimits_i |{\bm D}_{\perp}\dq[i]|^2 + \kappa_2 \left[ \phantom{~~~}\sum_{\makebox[0pt]{$\scriptstyle \bm{Q}_j = \pm \bm{Q}_{1,2}$}} \phantom{} \left( |\Dxdq[j]|^2 - |\Dydq[j]|^2 \right) \phantom{~~} - \phantom{\bm{Q}} \sum_{\makebox[0pt]{$\scriptstyle \bm{Q}_k = \pm \bm{Q}_{3,4}$}} \phantom{} \left( |\Dxdq[k]|^2 - |\Dydq[k]|^2\right) \right] \\
     \shoveright{ + \kappa_3 \left[\phantom{~~~}\sum_{\makebox[0pt]{$\scriptstyle \bm{Q}_l = \pm \bm{Q}_{1,4}$}} \phantom{} [(\Dxdq[l])\Cydq[l] + c.c.] \phantom{~~} - \phantom{\bm{Q}} \sum_{\makebox[0pt]{$\scriptstyle \bm{Q}_m = \pm \bm{Q}_{2,3}$}} \phantom{} [(\Dxdq[m])\Cydq[m] + c.c.]\right]}\\
      + \kappa_4 \sum\nolimits_{i} |\Dzdq[i]|^2 + \frac{1}{2}(\bm{\nabla} \times \bm{A})^2,
\end{multline}
\end{widetext}
where ${\bm D}=-i\nabla- 2e {\bm A}$, ${\bm D}_{\perp}=(D_x,D_y)$, and ${\bm B}=\bm{\nabla}\times{\bm A}$. 
In the spatially homogeneous case (for which spatial variations of the order parameter are ignored), the possible ground states depend upon nine unknown
phenomenological constants.  This parameter space is too large to carry out a complete analysis of all the possible ground states. However, with the above action,
it is straightforward to find the conditions under which a particular state is  a local minimum. In the following, we therefore consider a simplified theory that
applies to a material with orthorhombic symmetry (such as YBCO). For this orthorhombic theory, a complete analysis can be carried out. This analysis yields a PDW state
that is compatible with experiment, this state we generalize to tetragonal symmetry. Prior to the discussion of the solvable orthorhombic theory,
we first consider the secondary order parameters that are relevant for PDW order.
%
\vspace{2mm}
\section{\label{sec: secondary} Secondary order parameters}
Different PDW ground states are distinguished by the secondary order parameters that are induced by the PDW order. These secondary order parameters play a central role
in situations in which the original PDW order does not appear either due to impurities or due to fluctuations. In some circumstance, these secondary order parameters
have also been named vestigial order \cite{fra14}. These secondary order parameters are identified by examining all possible bi-linear products
of the $\dq[i]$. This leads to five distinct kinds of secondary order: CDW \cite{agt08,ber09}, orbital density wave order (ODW) \cite{agt08} (with spatially
modulated orbital currents), translational invariant charge-4 superconductivity (4SC) \cite{ber09-2,rad09} (we do not consider finite-momentum charge-4
superconductivity), strain \cite{ber09-2,rad09}, and translational invariant loop current (LC) order. Specifically, the CDW order is given by
$\rho_{2{\bm Q}}\propto (\dq \ndq^* + \ndq \dq^*)$ or $\rho_{{\bm Q}_1-{\bm Q}_2} \propto (\dq[1] \dq[2]^* + \ndq[2] \ndq[1]^*)$, the ODW
order is given by $L^z_{{\bm Q}_1-{\bm Q}_2} \propto i(\dq[1] \ndq[2]^* - \dq[2] \ndq[1]^*)$, the 4SC order is given by $\Delta_4\propto \dq \ndq$, strain order is 
given by $\epsilon_i \propto (|\dq[1]|^2 + |\ndq[1]|^2 - |\dq[2]|^2 -|\ndq[2]|^2)$ \cite{ber09,rad09}, and the loop current order, which was discussed above, 
by $l_i \propto (|\dq[i]|^2-|\ndq[i]|^2)$.
\vspace{6mm}
\section{\label{sec: GLW-ortho} GLW action: orthorhombic symmetry}
Here we consider the orthorhombic variant of Fig.~\ref{fig: PDW-momenta}. The GLW action in this case allows all possible ground states to be found and further allows
for a analysis of preemptive loop current order discussed in the next section. The order parameter has four complex degrees of freedom and is represented by the
momenta given by the solid arrows in Fig.~\ref{fig: PDW-momenta}. The same symmetry considerations as above lead to the partition function
$Z\propto \int \Pi_i \mathcal{D} \Delta_i e^{-S_{0}}$ with GLW action $S_{0}$ given by
\begin{widetext}
\begin{multline}
S_0 = r_0\sum\nolimits_{i}|\dq[i]|^2  + \frac{\beta_1}{2}  \left (\sum\nolimits_{i}|\dq[i]|^2 \right)^2 + \frac{\beta_2}{2} \left( |\dq[1]|^2+|\ndq[1]|^2 -|\dq[2]|^2-|\ndq[2]|^2 \right)^2 \\
\label{free} \shoveright{+\frac{\beta_3}{2} \left( |\dq[1]|^2-|\ndq[1]|^2 -|\dq[2]|^2+|\ndq[2]|^2 \right)^2
+\frac{\beta_4}{2} \left(|\dq[1]|^2-|\ndq[1]|^2  + |\dq[2]|^2 - |\ndq[2]|^2 \right)^2} \\
\shoveright{+ \beta_5 \left[ \dq[1]\ndq[1](\dq[2]\ndq[2])^*  +\dq[2]\ndq[2](\dq[1]\ndq[1])^* \right]
+\kappa_1\sum\nolimits_i|{\bm D}_{\perp} \Delta_i|^2
+ \kappa_2\sum\nolimits_i \left(|D_x\Delta_i|^2-|D_y\Delta_i|^2\right)}\\
\shoveright{+ \kappa_3 \left[\left((D_x \dq[1])(D_y\dq[1])^* + (D_x \ndq[1])(D_y\ndq[1])^* - (D_x \dq[2])(D_y\dq[2])^*-(D_x \ndq[2])(D_y\ndq[2])^* \right) + c.c.\right]} \\
+ \kappa_4\sum\nolimits_i|\Dzdq[i]|^2+\frac{1}{2}(\bm{\nabla} \times \bm{A})^2.
\end{multline}
\end{widetext}
\begin{table*}[t]
\caption{\label{tbl: ortho}{\bf Properties of PDW Ground States for orthorhombic symmetry in Fig.~\ref{fig: PDW-momenta}.} All possible PDW ground states and
accompanying CDW and ODW order. The second column shows the parameter regions for which these phases are stable. In the third and fourth columns: $2{\bm Q}_x=(2Q, 0)$,
$2{\bm Q}_y=(0, 2{Q})$, other modes can be found by using the relationships $\rho_{\bm Q}=(\rho_{-{\bm Q}})^*$ and $L^z_{\bm Q}=(L^z_{-{\bm Q}})^*$. The fifth
column gives all translational invariant order parameters with $l_x\propto$ $ |\dq[1]|^2-|\ndq[1]|^2-|\dq[2]|^2+|\ndq[2]|^2$,
$l_y\propto |\dq[1]|^2-|\ndq[1]|^2+|\dq[2]|^2-|\ndq[2]|^2$, $\Delta_{4e,s}\propto \dq[1]\ndq[1]+ \dq[2]\ndq[2]$, $\Delta_{4e,d}\propto \dq[1]\ndq[1]-\dq[2]\ndq[2]$, and
$\epsilon_{xy}\propto |\dq[1]|^2+|\ndq[1]|^2-|\dq[2]|^2-\ndq[2]|^2$. The sixth column gives the degeneracy of the ground state.}
\begin{ruledtabular}
\begin{tabular}{c r c c c r}
  $(\dq[1],\dq[2],\ndq[1],\ndq[2])$& Stability & CDW modes & ODW modes & Q=0 Order  &Degeneracy Manifold \\\hline
  \multirow{3}*{$(1,~0,~0,~0)$}	& $\beta_2+\beta_3<0,\beta_2+\beta_4<0$	& \multirow{3}*{none}			& \multirow{3}*{none}		& \multirow{3}*{\begin{tabular}{@{}c@{}}$\epsilon_{xy}$ \\ $l_x$, $l_y$\end{tabular}}	& \multirow{3}*{$U(1)\times Z_2\times Z_2$}	\\
				& $\beta_3+\beta_4<0$			&					&				& 		&				\\
				& $\beta_2+\beta_3+\beta_4<-|\beta_5|/4$&					&				&			&				\\\cline{2-2}
  \multirow{2}*{$(1,~1,~0,~0)$}	& $\beta_2+\beta_3>0,\beta_4<\beta_2$ 	& \multirow{2}*{$\rhoqx$} 		& \multirow{2}*{$L^z_{2{\bm Q}_x}$}& \multirow{2}*{$l_y$}& \multirow{2}*{$U(1)\times U(1) \times Z_2$}	\\
				& $\beta_4<\beta_3,\beta_4<-|\beta_5|/4$&					&				&			&				\\\cline{2-2}
  \multirow{2}*{$(1,~0,~0,~1)$}	& $\beta_2+\beta_4>0,\beta_3<\beta_2$ 	& \multirow{2}*{$\rhoqy$}		& \multirow{2}*{$L^z_{2{\bm Q}_y}$}& \multirow{2}*{ $l_x$}& \multirow{2}*{$U(1)\times U(1) \times Z_2$}	\\
				& $\beta_3<\beta_4,\beta_3<-|\beta_5|/4$&					&				&			& 				\\\cline{2-2}
  \multirow{2}*{$(1,~0,~1,~0)$}	& $\beta_3+\beta_4>0,\beta_2<\beta_3$	& \multirow{2}*{$\rho_{2{\bm Q}_1}$}	& \multirow{2}*{none} 		&$\epsilon_{xy}$ 	& \multirow{2}*{$U(1)\times U(1)\times Z_2$}	\\
				& $\beta_2<\beta_4,\beta_2<-|\beta_5|/4$&					&				&$\Delta_{4e,s}$, $\Delta_{4e,d}$	&		\\\cline{2-2}
  \multirow{3}*{$(1,~1,~1,~1)$}	& $\beta_5<0,\beta_5<4\beta_2$ 		& \multirow{3}*{\begin{tabular}{@{}c@{}}$\rho_{2{\bm Q}_1},~\rho_{2{\bm Q}_2}$ \\ $\rhoqx,~\rhoqy$ \end{tabular}}	& \multirow{3}*{none}		& \multirow{3}*{$\Delta_{4e,s}$}	& \multirow{3}*{$U(1)\times U(1)\times U(1)$}	\\
				& $\beta_5<4\beta_3,\beta_5<4\beta_4$	& 					&				& 			&				\\
				& $\beta_5/4<\beta_2+\beta_3+\beta_4$	&					&				&			&				\\\cline{2-2}
  \multirow{3}*{$(1,~i,~1,~i)$}	& $\beta_5>0,-\beta_5<4\beta_2$ 	& \multirow{3}*{$\rho_{2{\bm Q}_1},~\rho_{2{\bm Q}_2}$} 	& \multirow{3}*{$L^z_{2{\bm Q}_y},L^z_{2{\bm Q}_x}$}	& \multirow{3}*{$\Delta_{4e,d}$}					& \multirow{3}*{$U(1)\times U(1)\times U(1)$}	\\
				&$-\beta_5<4\beta_3,-\beta_5<4\beta_4$	&					&				&			&		\\
				&$-\beta_5/4<\beta_2+\beta_3+\beta_4$	&					&				&			&		\\
\end{tabular}
\end{ruledtabular}
\end{table*}
\subsection{\label{subsec: ground}Ground states}
\begin{figure}[t]
\begin{center}
\includegraphics[width=3.0in]{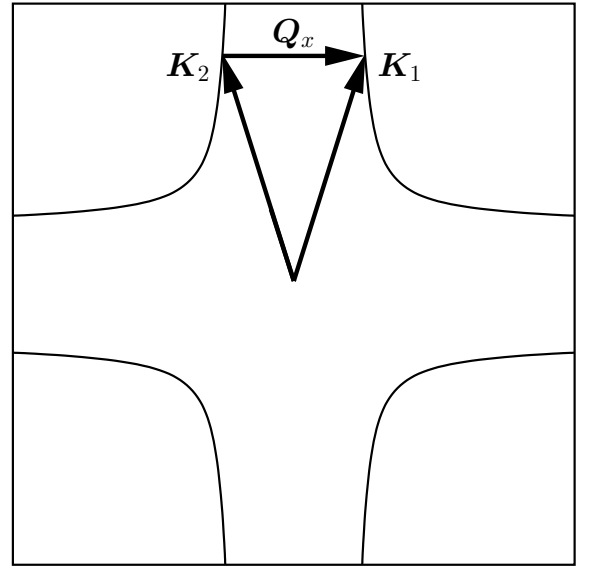}
\end{center}
\caption{The ME PDW state for orthorhombic symmetry. The arrows ${\bm K}_i$ depict the non-zero components of the PDW order parameter (which order at 
${\bm Q}_i=2{\bm K}_i$).  Together with the PDW order at the two wavevectors ${\bm Q}_i$, this state has CDW order at the wavevector $2{\bm Q}_x={\bm Q}_1-{\bm Q}_2$, 
ODW order at the same wavevector, and ME loop current order.}
\label{fig: varma}
\end{figure}

For this action, it is possible to find all homogeneous mean-field ground states analytically. These are listed in Table~\ref{tbl: ortho} together with the
corresponding conditions that the ground state represents a global minimum, secondary order parameters, and degeneracy manifold (degeneracy manifold specifies the
number of states with the same ground state energy). Of the ground states listed in Table \ref{tbl: ortho}, only one state (named the ME PDW state) has the potential to
represent a pseudogap mean-field order parameter when generalized to tetragonal symmetry. This ME PDW state has the order parameter
$(\dq[1],\dq[2],\ndq[1],\ndq[2])=\Delta(1,~1,~0,~0)$ and is depicted in Fig.~\ref{fig: varma}. It is stable when $\beta_1+\beta_2>0,~\beta_2+\beta_3>0,~\beta_4<\beta_2,
~\beta_4<\beta_3$, and $\beta_4<-|\beta_5|/4$. This state can be characterized by the secondary orders that it induces: loop current order $l_y=|\dq[1]|^2-|\ndq[1]|^2+
|\dq[2]|^2-|\ndq[2]|^2$; CDW order $\rho_{2{\bm Q}_x} = \dq[1]\dq[2]^*+\ndq[2]\ndq[1]^*$; and orbital density wave (ODW) order at the same wavevector as the CDW order
$L^z_{2{\bm Q}_x}=i(\dq[1]\ndq[2]^*-\dq[2]\ndq[1]^*)$ ($L^z$ is the $z$-component of angular momentum). The ground state manifold of the ME PDW state has a
$U(1)\times U(1)\times Z_2$ degeneracy. The two $U(1)$ degeneracies arise from the usual SC phase symmetry breaking and from the breaking of translational invariance.
The $Z_2$ symmetry denotes the degeneracy between the $(\dq[1],\dq[2],\ndq[1],\ndq[2])=\Delta(1,~1,~0,~0)$ and $\Delta(0,~0,~1,~1)$ states and is associated with the
ME loop current order (which is of opposite sign for these two degenerate states). In the next section we discuss how this ground state manifold can give rise to a
preemptive transition for which there is only ME loop current long-range order.
\subsection{\label{subsec: preemp-ortho} Emergent loop current order - Orthorhombic symmetry }
Fluctuations can lead to a preemptive transition in which the $U(1)\times U(1)$ symmetry is not broken, but the $Z_2$ symmetry is. Such a state will exhibit spatial
long-range ME loop current order and short-range SC and CDW order. To examine this possibility, we consider the partition function given by the effective action in 
Eq.~(\ref{free}) in two dimensions (2D), ignore the vector potential, and focus on the parameter regime for which the ME PDW state is stable. We decouple the quartic
terms through Hubbard-Stratonovich (HS) transformations. In particular,  we introduce the field  $\psi$ to decouple the  $(\sum_i|\Delta_i|^2)^2$ term, $\epsilon_{xy}$ 
to decouple the $(\dq[1]|^2 + |\ndq[1]|^2-|\dq[2]|^2 - |\ndq[2]|^2)^2$ term, $l_x$ to decouple the $ (|\dq[1]|^2-|\ndq[1]|^2-|\dq[2]|^2+|\ndq[2]|^2)^2$ term, $l_y$ to 
decouple the $( |\dq[1]|^2-|\ndq[1]|^2+|\dq[2]|^2-|\ndq[2]|^2)^2$ term, and two complex fields $\Delta_{4e,s}$ and $\Delta_{4e,d}$ to decouple the
$\left[ \dq[1]\ndq[1](\dq[2]\ndq[2])^*  +\dq[2]\ndq[2](\dq[1]\ndq[1])^* \right ]$ term. The resultant action is quadratic in the fields $\dq[i]$ and these fields can
be integrated out.  For the parameter regime we examine, the phases with non-zero $\Delta_{4e,s}$ and $\Delta_{4e,d}$ are energetically unfavorable.  Consequently we
set these fields to zero. Additionally, the remaining fields have Ising symmetry, so it is reasonable to treat these at a mean-field level. This leads to the following
effective action
\setlength\multlinegap{0pt}
\begin{multline}
\frac{S_{\text{eff}}}{A}=\frac{l_x^2}{2|\beta_3|}+\frac{l_y^2}{2|\beta_4|}-\frac{\psi^2}{2\beta_1}-\frac{\epsilon_{xy}^2}{2\beta_2}\\
		  + \int \frac{d^2 q}{4 \pi^2} \ln \Big [(\chi_{1,\bm{q}}^{-1}+\epsilon_{xy}+l_x+l_y)(\chi_{1,\bm{q}}^{-1}+\epsilon_{xy}-l_x-l_y)\\
		  (\chi_{2,\bm{q}}^{-1}-\epsilon_{xy}+l_x-l_y)(\chi_{2,\bm{q}}^{-1}-\epsilon_{xy}-l_x+l_y)\Big],
\end{multline}
where $A$ is the area, $\chi_{1,\bm{q}}^{-1}=r_0+\psi+\kappa_1q^2+\kappa_2(q_x^2-q_y^2)+2\kappa_3 q_xq_y$,
$\chi_{2,\bm{q}}^{-1} = r_0+\psi+\kappa_1q^2+\kappa_2(q_x^2-q_y^2)-2\kappa_3 q_xq_y$.
The anisotropy due to $\kappa_2$ and $\kappa_3$ can be removed by rotating and re-scaling $q_x$ and $q_y$, yielding $(\tilde{q}_x^2+\tilde{q}_y^2)/ \tilde{\kappa}$
with $\tilde{\kappa}=\sqrt{\kappa_1^2-\kappa_2^2-\kappa_3^2}$, and the integrals over momenta can then be carried out. Treating $S_{\text{eff}}$ within a mean field
approximation leads to the following self-consistency equations
\begin{multline}
r^* = \bar{r}_0-\tilde{\beta}_1 \ln \big\{ \big[(r^*+\epsilon_{xy}^*)^2-(l_x^* + l_y^*)^2]\\
	      [(r^* -\epsilon_{xy}^*)^2-(l_x^* -l_y^*)^2 \big\rbrack \big\rbrace, \nonumber
\end{multline}
\begin{equation}
\begin{gathered} \nonumber
\epsilon_{xy}^* = - \tilde{\beta}_2 \ln\left [\frac{(r^*+\epsilon_{xy}^*)^2-(l_x^* + l_y^*)^2}{(r^*-\epsilon_{xy}^*)^2-(l_x^* - l_y^*)^2)}\right ],\\
l_x^* = - \tilde{\beta}_3\ln\left[\frac{(r^*+l_x^*)^2 - (\epsilon_{xy}^* + l_y^*)^2}{(r^* - l_x^*)^2 - (\epsilon_{xy}^* - l_y^*)^2}\right],\\
l_y^* = \ln\left[\frac{(r^*+l_y^*)^2 - (\epsilon_{xy}^* + l_x^*)^2}{(r^* - l_y^*)^2 - (\epsilon_{xy}^* - l_x^*)^2}\right],
\end{gathered}
\end{equation}
where $r^*=r_0^*+\psi^*$, the $^*$ denotes a rescaling by a factor
$4\pi\tilde{\kappa}/|\beta_4|$, $\tilde{\beta}_i=\beta_i/|\beta_4|$, $\bar{r}_0=r_0^*+8 \tilde{\beta}_1 \ln \Lambda + 4\bar{\beta_1}\ln(4\pi\tilde{\kappa}/|\beta_4|)$
and $\Lambda$ is the momentum cutoff. We find that for parameters $\beta_i$ such that the ME PDW state is stable, the mean field solution is given by 
$\epsilon_{xy}=l_x=0$ and $l_y\ne 0$. The mathematical analysis of this solution is the same as that used to examine preemptive nematic order in 
Ref.~\onlinecite{fer12}. This work implies that there is a second order transition into a ME loop current state when $\tilde{\beta}_1>2$ (this becomes first order 
transition  if $\tilde{\beta}_1<2$). This analysis can be extended to three dimensions and, provided $\kappa_4/\tilde{\kappa}$ is sufficiently small, a second order 
transition into a loop current phase will occur \cite{fer12}. Such a preemptive ME loop current phase will exhibit: SC and CDW correlations consistent with experiment 
\cite{ghi12,com13,sil13,li10,yu14}; broken time-reversal symmetry; broken parity symmetry; and is invariant under the product of time-reversal and parity symmetry.
\begin{figure}[t]
\begin{center}
\subfloat[]{\label{subfig: varma2}
\includegraphics[width=1.62in]{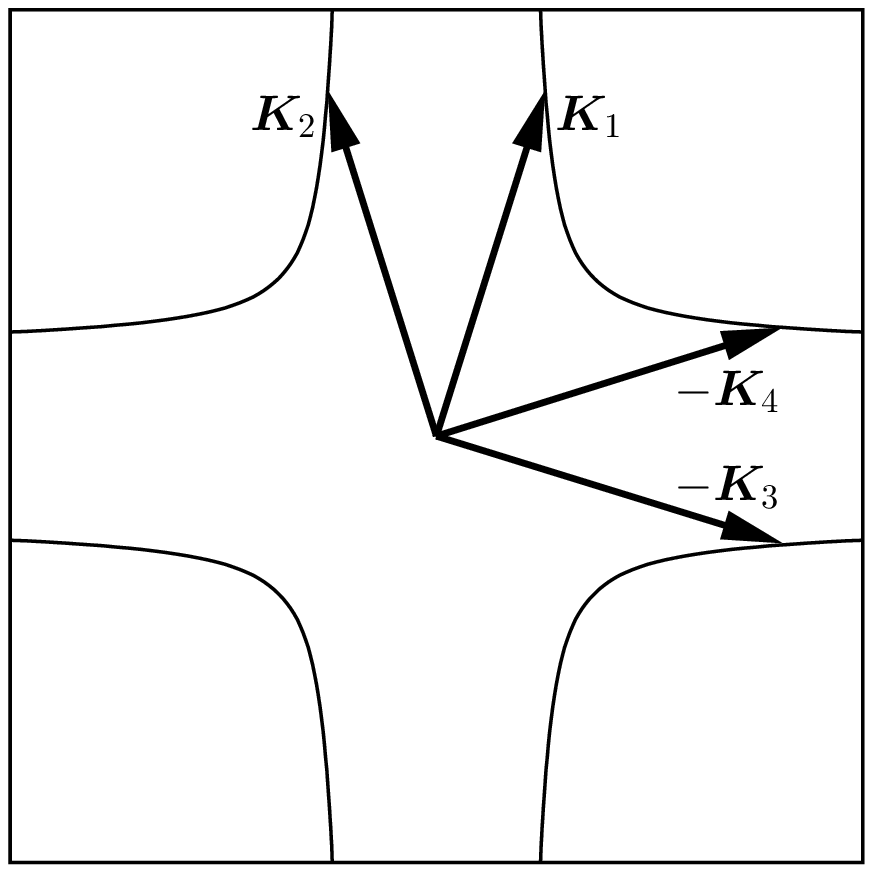}} \hfill
\subfloat[]{\label{subfig: varma3}
\includegraphics[width=1.62in]{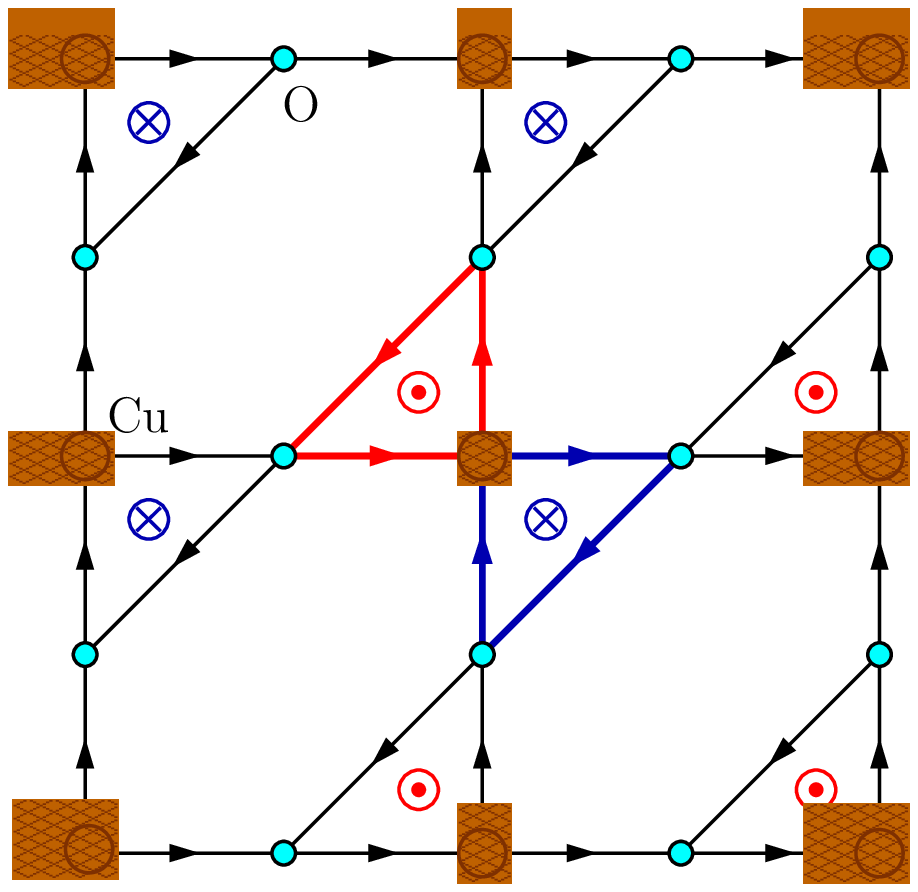}}
\caption{(Color online) The ME PDW state for tetragonal symmetry. (a)  The arrows ${\bm K}_i$ depict the non-zero components of the PDW order parameter in the ME PDW 
state (which order at ${\bm Q}_i=2{\bm K}_i$). This state has the same symmetry properties as the ME loop current phase discussed in Ref.~\onlinecite{sim02}.
(b) ME Loop current state introduced in Ref.~\onlinecite{sim02}. Here the larger dark circles are Cu sites, the smaller circles are O sites, the arrows represent the 
direction of the current, and the arrow heads and tails give the direction of the magnetic moments induced by the currents. \label{fig: var}}
\vspace{-5mm}
\end{center}
\end{figure}

\section{\label{sec: in-plane-tetra}In-plane loop current order - tetragonal symmetry}
\vspace{-2mm}
The ME PDW state found in Section~\ref{subsec: ground} has a natural generalization to tetragonal symmetry. In particular,
$(\dq[1], \dq[2], \dq[3], \dq[4], \ndq[1], \ndq[2], \ndq[3], \ndq[4]) = (\Delta_1, \Delta_2, 0, 0, 0, 0, \Delta_2, \Delta_1)$
is a stable state of the tetragonal GLW action (this will become apparent in the analysis that follows). This state is depicted in 
Fig.~\ref{fig: var}\subref{subfig: varma2}, it shares the same symmetries as the ME loop current state shown in Fig.~\ref{fig: var}\subref{subfig: varma3} which has been 
discussed in Refs.~\onlinecite{sim02, aji13}. Note that $\Delta_1 \ne \Delta_2$, however, as $\delta K_y = 0$, we recover the state examined in Ref.~\onlinecite{lee14} for which 
$\Delta_1=\Delta_2$, so for sufficiently small $\delta K_y$, we expect that $\Delta_1\approx \Delta_2$. To carry out an analysis of this phase, we follow the approach used 
in Section V for orthorhombic symmetry. In particular, we re-write the free energy terms denoted by $\beta_1$ to $\beta_7$ as squares of basis functions of irreducible 
invariants for tetragonal symmetry. This allows for a straightforward HS transformation. While we can also introduce HS fields for the terms $\beta_{ci}$, for the loop 
current phases we are interested in, these fields vanish (as they did in the orthorhombic case), consequently, we will not include these terms in the following. To 
reformulate the quartic portion of the effective action, we set $l_i = |\dq[i]|^2 - |\ndq[i]|^2$ and $ \epsilon_i = |\dq[i]|^2 + |\ndq[i]|^2 $. Basis functions for 
irreducible representations of $D_{4h}$ are then $p_{1x} = - l_3 - l_4$, $p_{1y} = l_1 + l_2$, $p_{2x} = l_1 - l_2$, $p_{2y} = l_3 - l_4$ ($\bm{p}_1$ and $\bm{p}_2$ are 
both bases for the $E_u$ representation), $\psi=\sum_i \epsilon_i$ (corresponding to the $A_{1g}$ representation), $\gamma = \epsilon_1-\epsilon_2+\epsilon_3-\epsilon_4$ 
(corresponding to the $A_{2g}$ representation), $\epsilon_{x^2-y^2}=\epsilon_1+\epsilon_2-\epsilon_3-\epsilon_4$ (corresponding to the $B_{1g}$ representation),
$\epsilon_{xy}=\epsilon_1-\epsilon_2-\epsilon_3+\epsilon_4$ (corresponding to the $B_{2g}$ representation). In terms of these basis functions Eq.~(\ref{eq: tetra-free1_hom}) 
can be rewritten as
\begin{widetext}
\begin{multline} \label{eq: tetra-free2_hom}
 S_{0, \text{hom}} = r_0 \sum\nolimits_i |\dq[i]|^2 + \tilde\beta_1 \psi^2 + \tilde\beta_2 \bm{p}_{1}^2 + \tilde\beta_3 \bm{p}_{2}^2 + \tilde\beta_4 \bm{p}_1 . \bm{p}_2
								    + \tilde\beta_5 \gamma^2 + \tilde\beta_6 \epsilon_{x^2 -y^2}^2 + \tilde\beta_7 \epsilon_{xy}^2\\
				    \phantom{ABCDE} + \beta_{c_1} \left\{ \left[\dq[1]\ndq[1](\dq[2]\ndq[2])^*  + \dq[3]\ndq[3](\dq[4]\ndq[4])^* \right] + c.c.  \right\}\\
				    \phantom{ABCDE} + \beta_{c_2} \left\{ \left[\dq[1]\ndq[1](\dq[3]\ndq[3])^*  + \dq[2]\ndq[2](\dq[4]\ndq[4])^* \right] + c.c.  \right\}\\
				    + \beta_{c_3} \left\{ \left[\dq[1]\ndq[1](\dq[4]\ndq[4])^*  + \dq[2]\ndq[2](\dq[3]\ndq[3])^* \right] + c.c.  \right\},
\end{multline}
where $\tilde\beta_1 = \beta_1 + (1/8)(\beta_4 + \beta_6 + \beta_7 - \beta_2)$, $ \tilde\beta_2 = (1/8)(\beta_3 - \beta_2 - \beta_7)$,
$\tilde\beta_3 = (1/8)(\beta_7 - \beta_2 - \beta_3)$, $\tilde\beta_4 = (1/4)(\beta_6 - \beta_5)$, $\tilde\beta_5 = (1/8)(\beta_4 - \beta_3 - \beta_6)$,
$\tilde\beta_6 = ({\beta_2}/4) + (1/8)(\beta_3 - \beta_4 - \beta_6)$, $\tilde\beta_7 = (1/8)(\beta_2 + \beta_6 - \beta_4 - \beta_7)$. In the above expression,
all terms except $\beta_{ci}$ and $\tilde{\beta}_4$ are squares of basis functions. To account for $\tilde{\beta}_4$, we rotate 
$l_{1i}=\cos\theta ~p_{1i}+\sin\theta ~p_{2i}$ and $l_{2i}= -\sin\theta ~p_{1i}+\cos\theta ~p_{2i}$ with
$\cos\theta = \frac{\sqrt{\left(\tilde\beta_2 - \tilde\beta_3 + \sqrt{(\tilde\beta_2 - \tilde\beta_3)^2 + {\tilde\beta_4}^2}\right)^2 + 
{\tilde\beta_4}^2}}{2\sqrt{(\tilde\beta_2 - \tilde\beta_3)^2 + {\tilde\beta_4}^2}}$
and
$\sin\theta = \frac{\sqrt{\left(\tilde\beta_2 - \tilde\beta_3 - \sqrt{(\tilde\beta_2 - \tilde\beta_3)^2 + {\tilde\beta_4}^2}\right)^2 + 
{\tilde\beta_4}^2}}{2\sqrt{(\tilde\beta_2 - \tilde\beta_3)^2 + {\tilde\beta_4}^2}}.$
%
In terms of these new parameters Eq.~(\ref{eq: tetra-free2_hom}) can be expressed as
($\tilde\beta $\textquotesingle s and $\lambda$\textquotesingle s have been rescaled by a factor of half for convenience) 
\begin{multline} \label{eq: tetra-free3_hom}
  S_{0,\text{hom}} = r_0 \sum\nolimits_i |\dq[i]|^2 + \frac{\tilde\beta_1}{2} \psi^2 + \frac{\lambda_1}{2} \left( l_{1x}^2 + l_{1y}^2 \right) + \frac{\lambda_2}{2} \left( l_{2x}^2 + l_{2y}^2 \right)
					+ \frac{\tilde\beta_5}{2} \gamma^2 + \frac{\tilde\beta_6}{2} \epsilon_{x^2 -y^2}^2 + \frac{\tilde\beta_7}{2} \epsilon_{xy}^2\\
				    \phantom{ABCD} + \beta_{c_1} \left\{ \left[\dq[1]\ndq[1](\dq[2]\ndq[2])^*  + \dq[3]\ndq[3](\dq[4]\ndq[4])^* \right] + c.c.  \right\}\\
				    \phantom{ABCD} + \beta_{c_2} \left\{ \left[\dq[1]\ndq[1](\dq[3]\ndq[3])^*  + \dq[2]\ndq[2](\dq[4]\ndq[4])^* \right] + c.c.  \right\}\\
				    \shoveright{ + \beta_{c_3} \left\{ \left[\dq[1]\ndq[1](\dq[4]\ndq[4])^*  + \dq[2]\ndq[2](\dq[3]\ndq[3])^* \right] + c.c.  \right\}} \\
      \shoveleft{\phantom{ABC}= r_0 \sum\nolimits_i |\dq[i]|^2 + \frac{\tilde\beta_1}{2}{\left( \sum\nolimits_i |\dq[i]|^2 \right)}^2} \\
      \phantom{~~~}+ \frac{\lambda_1}{2}\left[\left(- |\dq[3]|^2 + |\ndq[3]|^2  - |\dq[4]|^2 + |\ndq[4]|^2\right)\cos\theta + \left(|\dq[1]|^2 - |\ndq[1]|^2 - |\dq[2]|^2 + |\ndq[2]|^2\right)\sin\theta \right]^2\\
      \phantom{~~~}+ \frac{\lambda_1}{2}\left[\left(|\dq[1]|^2 - |\ndq[1]|^2  + |\dq[2]|^2 - |\ndq[2]|^2\right)\cos\theta + \left(|\dq[3]|^2 - |\ndq[3]|^2 - |\dq[4]|^2 + |\ndq[4]|^2\right)\sin\theta \right]^2 \phantom{~~} \\
      \phantom{~~~}+ \frac{\lambda_2}{2}\left[\left(|\dq[3]|^2 - |\ndq[3]|^2  + |\dq[4]|^2 - |\ndq[4]|^2\right)\sin\theta + \left(|\dq[1]|^2 - |\ndq[1]|^2 - |\dq[2]|^2 + |\ndq[2]|^2\right)\cos\theta \right]^2 \phantom{~~} \\
      \phantom{~~~}+ \frac{\lambda_2}{2}\left[\left(- |\dq[1]|^2 + |\ndq[1]|^2  - |\dq[2]|^2 + |\ndq[2]|^2\right)\sin\theta + \left(|\dq[3]|^2 - |\ndq[3]|^2 - |\dq[4]|^2 + |\ndq[4]|^2\right)\cos\theta \right]^2 \\
      \phantom{~~~}+ \frac{\tilde\beta_5}{2} \left( |\dq[1]|^2 + |\ndq[1]|^2 -|\dq[2]|^2 - |\ndq[2]|^2 + |\dq[3]|^2 + |\ndq[3]|^2 - |\dq[4]|^2 - |\ndq[4]|^2  \right)^2	\\
      \phantom{~~~}+ \frac{\tilde\beta_6}{2} \left( |\dq[1]|^2 + |\ndq[1]|^2 +|\dq[2]|^2 + |\ndq[2]|^2 - |\dq[3]|^2 - |\ndq[3]|^2 - |\dq[4]|^2 - |\ndq[4]|^2  \right)^2 	\\
      \phantom{~~~}+ \frac{\tilde\beta_7}{2} \left( |\dq[1]|^2 + |\ndq[1]|^2 -|\dq[2]|^2 - |\ndq[2]|^2 - |\dq[3]|^2 - |\ndq[3]|^2 + |\dq[4]|^2 + |\ndq[4]|^2  \right)^2 	\\
      \phantom{~~~~~~~.}+ \beta_{c_1} \left\{ \left[\dq[1]\ndq[1](\dq[2]\ndq[2])^*  + \dq[3]\ndq[3](\dq[4]\ndq[4])^* \right ] + c.c.  \right\}	\\
      \phantom{~~~~~~~.}+ \beta_{c_2} \left\{ \left[\dq[1]\ndq[1](\dq[3]\ndq[3])^*  + \dq[2]\ndq[2](\dq[4]\ndq[4])^* \right ] + c.c.  \right\}	\\
      + \beta_{c_3} \left\{ \left[\dq[1]\ndq[1](\dq[4]\ndq[4])^*  + \dq[2]\ndq[2](\dq[3]\ndq[3])^* \right ] + c.c.  \right\},
\end{multline}
where $\lambda_1 =\frac{\tilde\beta_2 + \tilde\beta_3 + \sqrt{\left(\tilde\beta_2 - \tilde\beta_3 \right)^2 + {\tilde\beta_4}^2}}{2}$ and 	$\lambda_2 = \frac{\tilde\beta_2 + \tilde\beta_3 - \sqrt{\left(\tilde\beta_2 - \tilde\beta_3 \right)^2 + {\tilde\beta_4}^2}}{2}$. Notice that if $\lambda_1<0$, $\beta_{ci}$ are sufficiently small, and all other quartic terms are positive, then the ME loop current phase will be the mean-field ground state. This is the limit that we will examine further. In particular, in the next paragraph, we examine preemptive loop current order emerging from this ME PDW phase.

We decouple the quartic terms of Eq.~(\ref{eq: tetra-free3_hom}) through HS transformations.  In particular, introducing $\psi$, $l_{1x}$, $l_{1y}$, $l_{2x}$, $l_{2y}$,
$\gamma$, $\epsilon_{x^2-y^2}$ and $\epsilon_{xy}$ to decouple the second($(\sum_i|\Delta_i|^2)^2$), third, fourth, fifth, sixth, seventh, eighth and ninth term
respectively. The resultant action is quadratic in the fields $\dq[i]$ and these fields can be integrated out.  As in the orthorhombic case, the terms with 
$\beta_{ci}$ do not contribute to the effective action in the ME PDW phase,  so we do not include these terms (the HS decomposition of these terms can proceed through 
charge-4e superconducting fields, ignoring these terms is equivalent to setting these fields to zero). The remaining fields have discrete symmetries, so it is 
reasonable to treat these at a mean-field level. This leads to the following effective action (note we have set $\lambda_1<0$ and all other quartic terms are positive)

\begin{multline} \label{eq: tetra-eff}
  \frac{S_{\text{eff},\text{tet}}}{A} =  \frac{l_{1x}^2 + l_{1y}^2}{2|\lambda_1|} - \frac{l_{2x}^2 + l_{2y}^2}{2 \lambda_2}
    - \frac{\psi^2}{2\tilde\beta_1} - \frac{\gamma^2}{2\tilde\beta_5} - \frac{\epsilon_{x^2 - y^2}^2}{2\tilde\beta_6} - \frac{\epsilon_{xy}^2}{2\tilde\beta_7} \\
    + \int \frac{d^2 q}{4 \pi^2} \ln \left[(\chi_{1,\bm{q}}^{-1} + \gamma + \epsilon_{x^2 - y^2} + \epsilon_{xy} - l_{1x}\sin\theta - l_{1y}\cos\theta
 			  + l_{2x}\cos\theta - l_{2y}\sin\theta) \right.\\
	    \left. \phantom{ABCDEFG}(\chi_{1,\bm{q}}^{-1} + \gamma + \epsilon_{x^2 - y^2} + \epsilon_{xy} + l_{1x}\sin\theta + l_{1y}\cos\theta
  			  - l_{2x}\cos\theta + l_{2y}\sin\theta) \right.\\
 	    \left. \phantom{ABCDEFG}(\chi_{2,\bm{q}}^{-1} - \gamma + \epsilon_{x^2 - y^2} - \epsilon_{xy} + l_{1x}\sin\theta - l_{1y}\cos\theta
 			  - l_{2x}\cos\theta - l_{2y}\sin\theta) \right.\\
 	    \left. \phantom{ABCDEFG}(\chi_{2,\bm{q}}^{-1} - \gamma + \epsilon_{x^2 - y^2} - \epsilon_{xy} - l_{1x}\sin\theta + l_{1y}\cos\theta
 			  + l_{2x}\cos\theta + l_{2y}\sin\theta) \right.\\
 	    \left. \phantom{ABCDEFG}(\chi_{3,\bm{q}}^{-1} + \gamma - \epsilon_{x^2 - y^2} - \epsilon_{xy} + l_{1x}\cos\theta - l_{1y}\sin\theta
 			  + l_{2x}\sin\theta + l_{2y}\cos\theta)\right.\\
 	    \left. \phantom{ABCDEFG}(\chi_{3,\bm{q}}^{-1} + \gamma - \epsilon_{x^2 - y^2} - \epsilon_{xy} - l_{1x}\cos\theta + l_{1y}\sin\theta
 			  - l_{2x}\sin\theta - l_{2y}\cos\theta)\right.\\
 	    \left. \phantom{ABCDEFG}(\chi_{4,\bm{q}}^{-1} - \gamma - \epsilon_{x^2 - y^2} + \epsilon_{xy} + l_{1x}\cos\theta + l_{1y}\sin\theta
 			  + l_{2x}\sin\theta - l_{2y}\cos\theta)\right.\\
 	    \left. \phantom{ABCDEFG}(\chi_{4,\bm{q}}^{-1} - \gamma - \epsilon_{x^2 - y^2} + \epsilon_{xy} - l_{1x}\cos\theta - l_{1y}\sin\theta
 			  - l_{2x}\sin\theta + l_{2y}\cos\theta)\right], 			
\end{multline}
where $\chi_{1,\bm{q}}^{-1} = r_0 + \psi + \kappa_1(q_x^2 + q_y^2) + \kappa_2(q_x^2 - q_y^2) + 2\kappa_3q_xq_y$, 
      $\chi_{2,\bm{q}}^{-1} = r_0 + \psi + \kappa_1(q_x^2 + q_y^2) + \kappa_2(q_x^2 - q_y^2) - 2\kappa_3q_xq_y$, 
$\chi_{3,\bm{q}}^{-1} = r_0 + \psi + \kappa_1(q_x^2 + q_y^2) - \kappa_2(q_x^2 - q_y^2) - 2\kappa_3q_xq_y$, and 
      $\chi_{4,\bm{q}}^{-1} = r_0 + \psi + \kappa_1(q_x^2 + q_y^2) - \kappa_2(q_x^2 - q_y^2) + 2\kappa_3q_xq_y$.

To carry out the integrals, the anisotropy in $\chi_{i, \bm q}^{-1}$ due to $\kappa_2$ and $\kappa_3$, can again be removed by rotating and re-scaling $q_x$ and $q_y$, 
yielding $(\tilde{q}_x^2+\tilde{q}_y^2)/ \tilde{\kappa}$ with $\tilde{\kappa}=\sqrt{\kappa_1^2-\kappa_2^2-\kappa_3^2}$. We find the self-consistency equations by 
setting the first derivatives with respect to the field equal to zero. The relevant solution that minimizes the action satisfies $\gamma = 0$, $\epsilon_{x^2-y^2} = 0$, 
$l_{1x} = l_{1y} \equiv \ell_1$ and $l_{2x} = l_{2y} \equiv \ell_2$ and the self consistency equations become (here $r = r_0 + \psi$ and 
$\bar{r}_0 = r_0 + (4\tilde\beta_1/\pi \tilde\kappa)\ln\Lambda$)
\begin{multline}\label{eq: r2}
 r = \bar{r}_0 - \frac{8\tilde\beta_1}{\pi \tilde\kappa} \ln \left\{ \left[ \left( r + \epsilon_{xy} \right)^2 - \left( \ell_1\cos\theta + \ell_1\sin\theta + \ell_2\sin\theta - \ell_2\cos\theta \right)^2 \right] \right. \\
							      \left. \left[ \left( r - \epsilon_{xy} \right)^2 - \left( \ell_1\cos\theta - \ell_1\sin\theta + \ell_2\sin\theta + \ell_2\cos\theta \right)^2 \right] \right\},
\end{multline}

\begin{multline}
 \epsilon_{xy} = - \frac{\tilde\beta_7}{4\pi\tilde\kappa} \left\{ \ln \left[ \frac{\left( r + \epsilon_{xy} \right)^2 - \left( \ell_1\sin\theta + \ell_1\cos\theta - \ell_2\cos\theta + \ell_2\sin\theta \right)^2}
										  {\left( r - \epsilon_{xy} \right)^2 - \left( \ell_1\cos\theta - \ell_1\sin\theta + \ell_2\sin\theta + \ell_2\cos\theta \right)^2}\right] \right. \\
							 \left. + \ln \left[ \frac{\left( r + \epsilon_{xy} \right)^2 - \left( \ell_1\cos\theta + \ell_1\sin\theta + \ell_2\sin\theta - \ell_2\cos\theta \right)^2}
										  {\left( r - \epsilon_{xy} \right)^2 - \left( \ell_1\sin\theta - \ell_1\cos\theta - \ell_2\cos\theta - \ell_2\sin\theta \right)^2}\right] \right\},
\end{multline}

\begin{multline}
\ell_1 = \frac{|\lambda_1|}{4\pi\tilde\kappa} \left\{ \cos\theta \ln \left[ \frac{\left( r + \ell_1\cos\theta + \ell_2\sin\theta \right)^2 - \left( \epsilon_{xy} + \ell_1\sin\theta - \ell_2\cos\theta \right)^2}
									      {\left( r - \ell_1\cos\theta - \ell_2\sin\theta \right)^2 - \left( \epsilon_{xy} - \ell_1\sin\theta + \ell_2\cos\theta \right)^2} \right] \right.\\
					  \left. + \sin\theta \ln \left[ \frac{\left( r + \ell_1\sin\theta - \ell_2\cos\theta \right)^2 - \left( \epsilon_{xy} + \ell_1\cos\theta + \ell_2\sin\theta \right)^2}
									      {\left( r - \ell_1\sin\theta + \ell_2\cos\theta \right)^2 - \left( \epsilon_{xy} - \ell_1\cos\theta - \ell_2\sin\theta \right)^2} \right] \right\},
\end{multline}

\begin{multline}
\ell_2 = -\frac{\lambda_2}{4\pi\tilde\kappa} \left\{ \cos\theta \ln \left[ \frac{\left( r - \ell_1\sin\theta + \ell_2\cos\theta \right)^2 - \left( \epsilon_{xy} - \ell_1\cos\theta - \ell_2\sin\theta \right)^2}
									     {\left( r + \ell_1\sin\theta - \ell_2\cos\theta \right)^2 - \left( \epsilon_{xy} + \ell_1\cos\theta + \ell_2\sin\theta \right)^2} \right] \right.\\
					 \left. + \sin\theta \ln \left[ \frac{\left( r + \ell_1\cos\theta + \ell_2\sin\theta \right)^2 - \left( \epsilon_{xy} + \ell_1\sin\theta - \ell_2\cos\theta \right)^2}
									     {\left( r - \ell_1\cos\theta - \ell_2\sin\theta \right)^2 - \left( \epsilon_{xy} - \ell_1\sin\theta + \ell_2\cos\theta \right)^2} \right] \right\}.
\end{multline}
\end{widetext}

To address whether or not there can be a second order transition into a phase with loop current order, we expand in powers of $\ell_1$. To cubic order in $\ell_1$ we 
find

\begin{multline}
 \epsilon_{xy} = - \frac{\tilde \beta_7^*}{2(2 \tilde \beta_7^* + r) r} \left[ 4 \cos2\theta ~ \ell_1 \ell_2 \right.\\  \left. + 2 \sin2\theta (-\ell_1^2 + \ell_2^2)\right],
\end{multline}
%
 \begin{equation}
   \ell_2 \sim \mathcal{O}(\ell_1^3), 
\end{equation}
\begin{multline}
 4 r^2(r - \left|\lambda_1^* \right|) \ell_1 = - 4\left|\lambda_1^* \right| \frac{\tilde \beta_7^*}{2(2 \tilde \beta_7^* + r)} \sin^22\theta~\ell_1^3 \\  - \frac{2}{3}\left|\lambda_1^* \right| \left( \cos4\theta - 3\right) \ell_1^3,
\end{multline}
where $^*$ denotes that the coefficients are scaled by $\pi \tilde\kappa$. Thus to leading order in $\ell_1$, $r = \left|\lambda_1^* \right|$. Going to next higher 
order, let $r = r_{\delta = 0} + \delta = \left|\lambda_1^* \right| + \delta$ where $\delta$ is small correction such that $(\delta/\left|\lambda_1^* \right| \ll 1)$, 
then the previous equation becomes
\begin{multline}\label{eq: delta}
 \frac{\delta}{\left|\lambda_1^* \right|} = \left(- \frac{\alpha_7}{2 \alpha_7 + 1} + \frac{1}{6} \right) \sin^22\theta~{\ell_1^*}^2 - \frac{1}{6}\cos^22\theta~{\ell_1^*}^2 \\ + \frac{1}{2} {\ell_1^*}^2
\end{multline}
and Eq.~(\ref{eq: r2}) leads to
\begin{equation}\label{eq: rbarbar}
 \bar{\bar{r}}_0 = 1 + \left(1 + 32 \alpha_1 \right) \frac{\delta}{\left|\lambda_1^* \right|} - 16 \alpha_1 {\ell_1^*}^2,
\end{equation}
where $\bar{\bar{r}}_0 = (\bar r_0 / \left|\lambda_1^* \right|) - 32 \alpha_1 \ln\left|\lambda_1^* \right| $, $\alpha_1 = \tilde \beta_1 / \left| \lambda_1 \right|$,
$ \alpha_7 = \tilde \beta_7 / \left| \lambda_1 \right| $ and $\ell_1^* = \ell_1 / \left| \lambda_1^* \right|$. Eliminating  $\delta$ between Eqs.~(\ref{eq: delta}) and
(\ref{eq: rbarbar}), we obtain
\begin{multline}\label{eq: trans}
 \bar{\bar{r}}_0 = 1 + (1 + 32 \alpha_1) \left[ \left(-\frac{\alpha_7}{1 + 2 \alpha_7} + \frac{1}{6} \right)\sin^22\theta \right.\\ \left. - \frac{1}{6}\cos^22\theta \right]{\ell_1^*}^2 + \frac{1}{2} {\ell_1^*}^2.
\end{multline}
Equation~(\ref{eq: trans}) shows that a local maximum $\bar{\bar{r}}_0 = 1$ occurs if the quadratic term in $\ell_1^*$ is negative. Since $\bar{\bar{r}}_0$ is monotonically
increasing with temperature, this maximum gives the highest possible transition temperature (provided there are no other local maxima at higher $\bar{\bar{r}}_0$ -- here
we note that no such maxima occurred in a related model \cite{fer12}) and the corresponding transition is second order. However, if the quadratic term in $\ell_1^*$ is
positive, then the largest value of $\bar{\bar{r}}_0$ will occur at non-zero $\ell_1^*$, indicating a first order transition. This emergent loop current phase shares
the same symmetry properties as the ME loop current state discussed in Refs.~\onlinecite{sim02, aji13}. While such a phase captures much of the physics associated with broken
time-reversal symmetry, it does not provide a complete explanation of all the signatures of broken time-reversal symmetry in the pseudogap phase \cite{yak14}. We 
address this in the next section. 
%
%

\section{\label{sec: tilted-loop}Tilted loop current order}
\begin{figure}[h]
\begin{center}
\subfloat[]{\label{subfig: yakovenko_loop}
\includegraphics[width=1.62in]{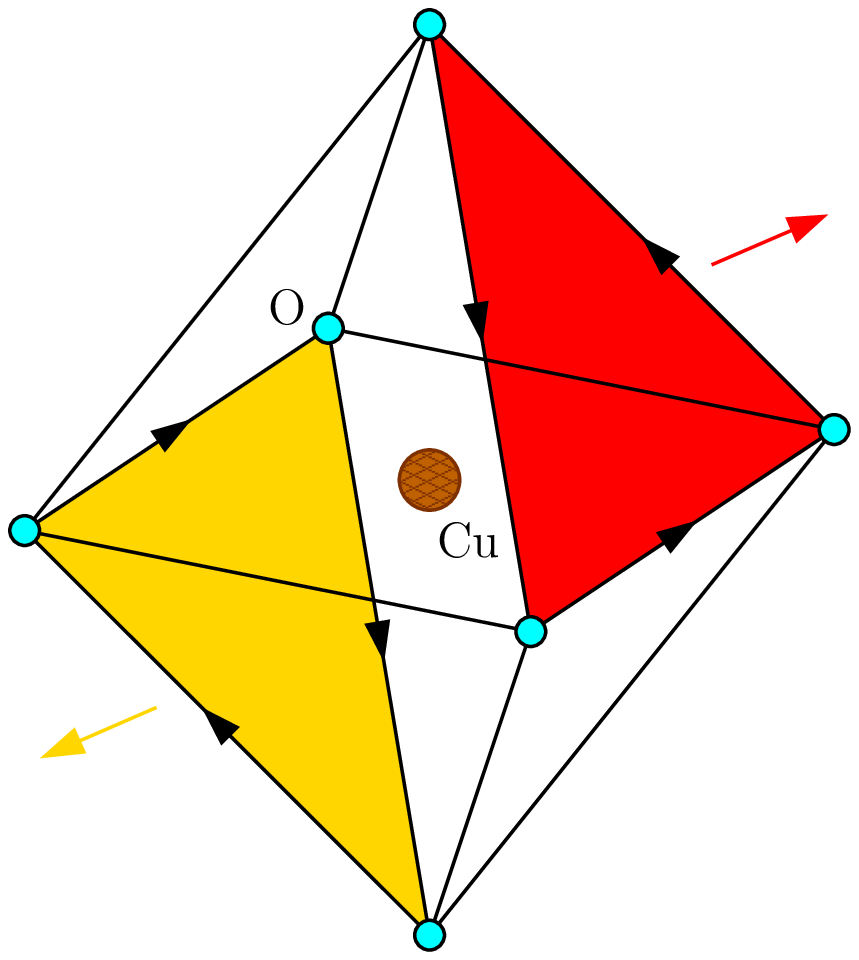}} \hfill
\subfloat[]{\label{subfig: yakovenko_pairing}
\includegraphics[width=1.62in]{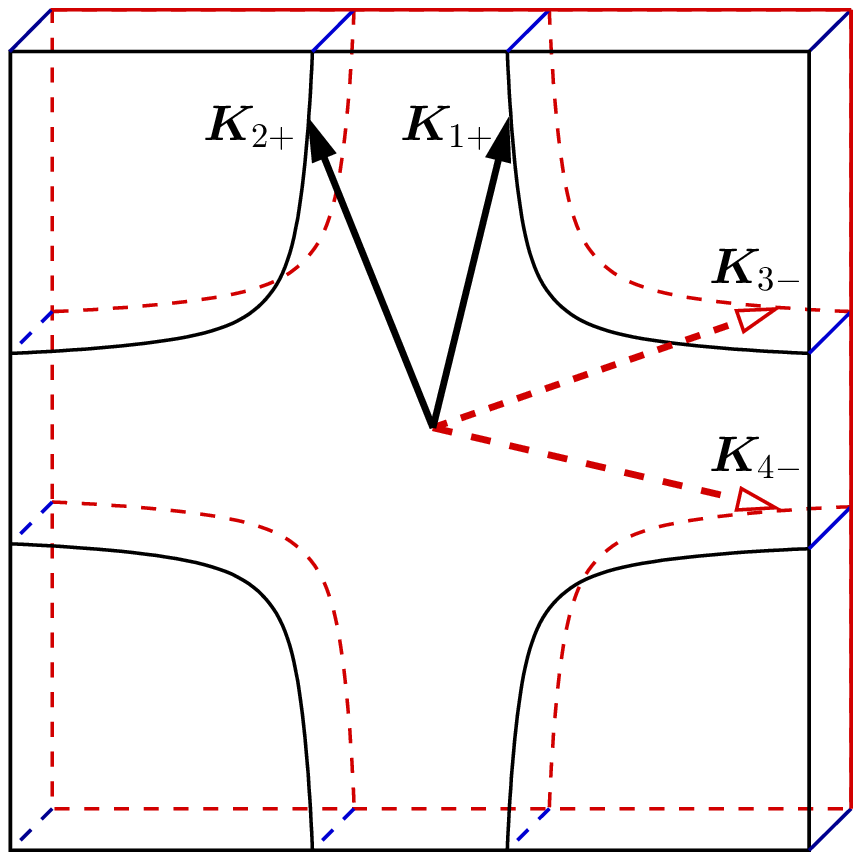}}
\caption{(Color online) (a) Tilted loop current state proposed by Yakovenko \cite{yak14}. The arrows on the bonds depict the direction of the current, the longer arrows 
depict the associated magnetic moments.
(b) PDW state with the same symmetry properties as the tilted loop current state. The arrows ${\bm K}_i$ depict the non-zero components of the PDW order parameter. 
Wavevectors labeled \textquotedblleft+\textquotedblright are above the $x$-$y$ plane and those labeled \textquotedblleft\textendash\textquotedblright are below the $x$-$y$ 
plane. \label{fig: yako}}
\vspace{-5mm}
\end{center}
\end{figure}

It has been argued that the Kerr effect \cite{xia08,kar14} is zero for the ME loop current state discussed above and a non-vanishing Kerr effect requires additional 
physics (such as a structural transition \cite{she13} or ordering along the $c$-axis). This has been discussed in detail by Yakovenko \cite{yak14} and he has 
identified a modified loop current state consistent with all experiments of broken time-reversal symmetry. This tilted loop current state is shown in 
Fig.~\ref{fig: yako}\subref{subfig: yakovenko_loop}. 
It is possible to find a PDW state that shares the same symmetry properties as the tilted loop current state (once the SC and CDW orders 
are removed through fluctuations). The simplest way to find such a state is to allow for the pairing momenta to have a $c$-axis component. The corresponding PDW order 
parameter has sixteen complex degrees of freedom (eight for momenta $\vQ_i+Q_z\hat{z}$ and eight for PDW momenta $\vQ_i-Q_z\hat{z}$ where the $\vQ_i$ are the momenta 
considered in Section \ref{sec: in-plane-tetra}). Here we do not present a complete analysis of this order parameter. However, it is possible to show that the state 
depicted in Fig.~\ref{fig: yako}\subref{subfig: yakovenko_pairing} is a mean-field ground state and thus represents a viable order parameter. In this state only four of 
the PDW momenta have non-zero order parameter components. As depicted in Fig.~\ref{fig: yako}\subref{subfig: yakovenko_pairing}, two of these momenta lie below the $x$-$y$ 
plane and two lie above the $x$-$y$ plane. When the SC  and CDW order are removed through fluctuations, this state will have the same symmetry properties as the tilted 
loop-current phase and is therefore also consistent with all existing experiments that show broken time-reversal symmetry.
\vspace{-2mm}
\section{\label{sec: quasi} Quasi-particle properties of loop current PDW phases}
%
In this Section we examine whether the broken time-
\newpage

\onecolumngrid

\begin{figure}[t]
\centering
\subfloat[$\Delta_1 = \Delta_2;~k_y = \pi$]{\label{subfig: banda}
\includegraphics[width=0.29\textwidth]{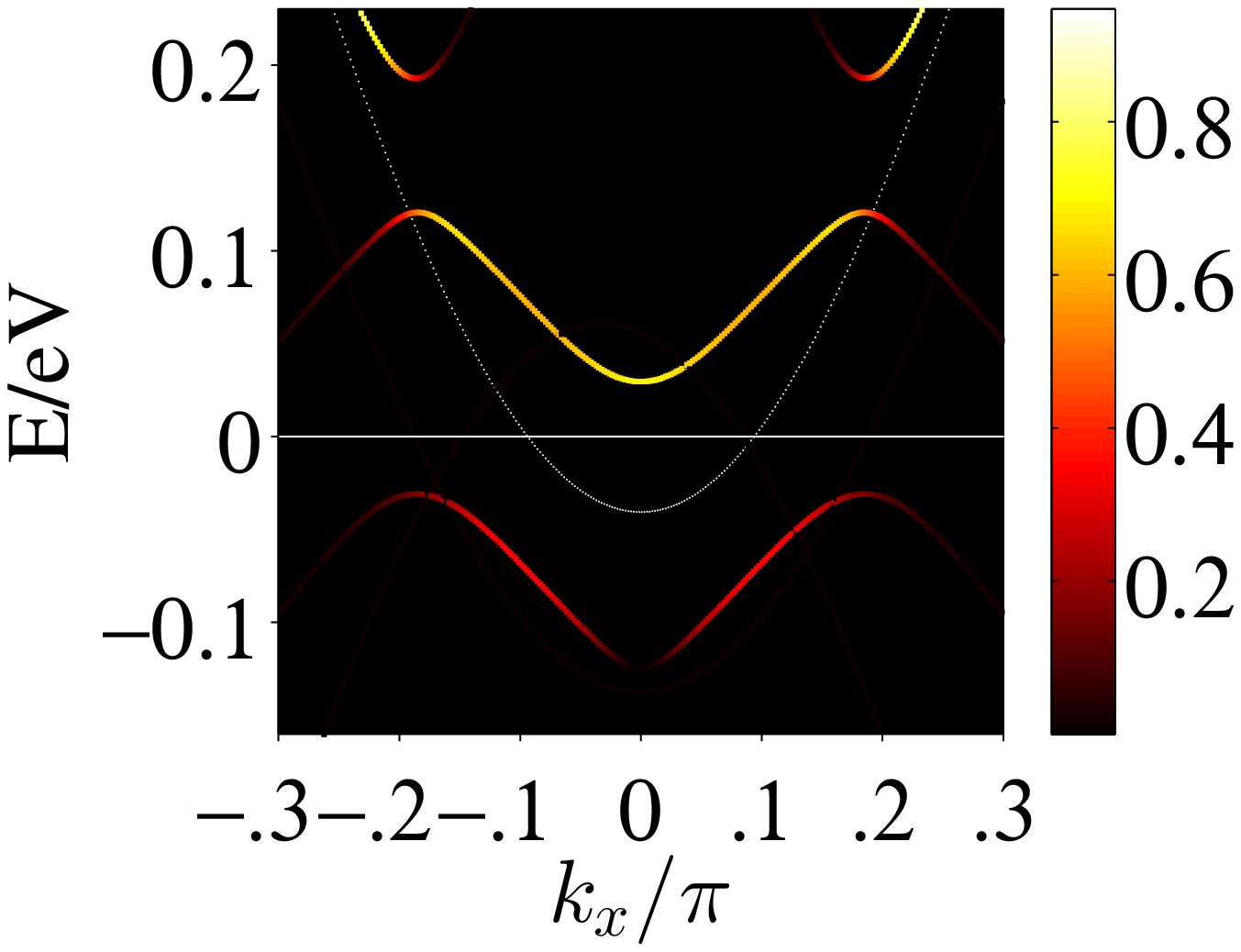}}
\subfloat[$\Delta_1 = \Delta_2;~k_y = \pi - 0.7$]{\label{subfig: bandb}
\includegraphics[width=0.29\textwidth]{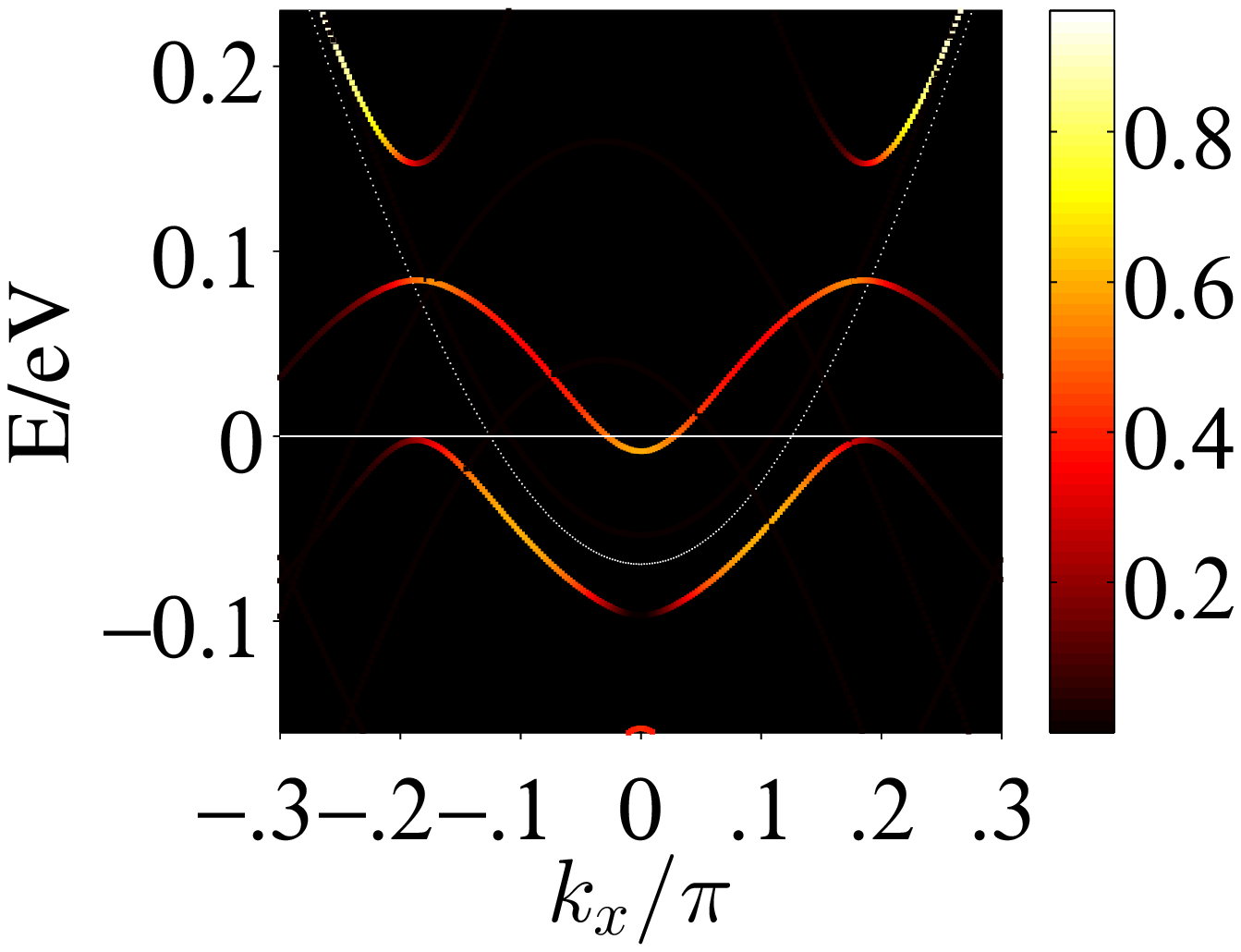}}
\subfloat[$\Delta_1 \ne \Delta_2;~k_y = \pi$]{\label{subfig: bandc}
\includegraphics[width=0.29\textwidth]{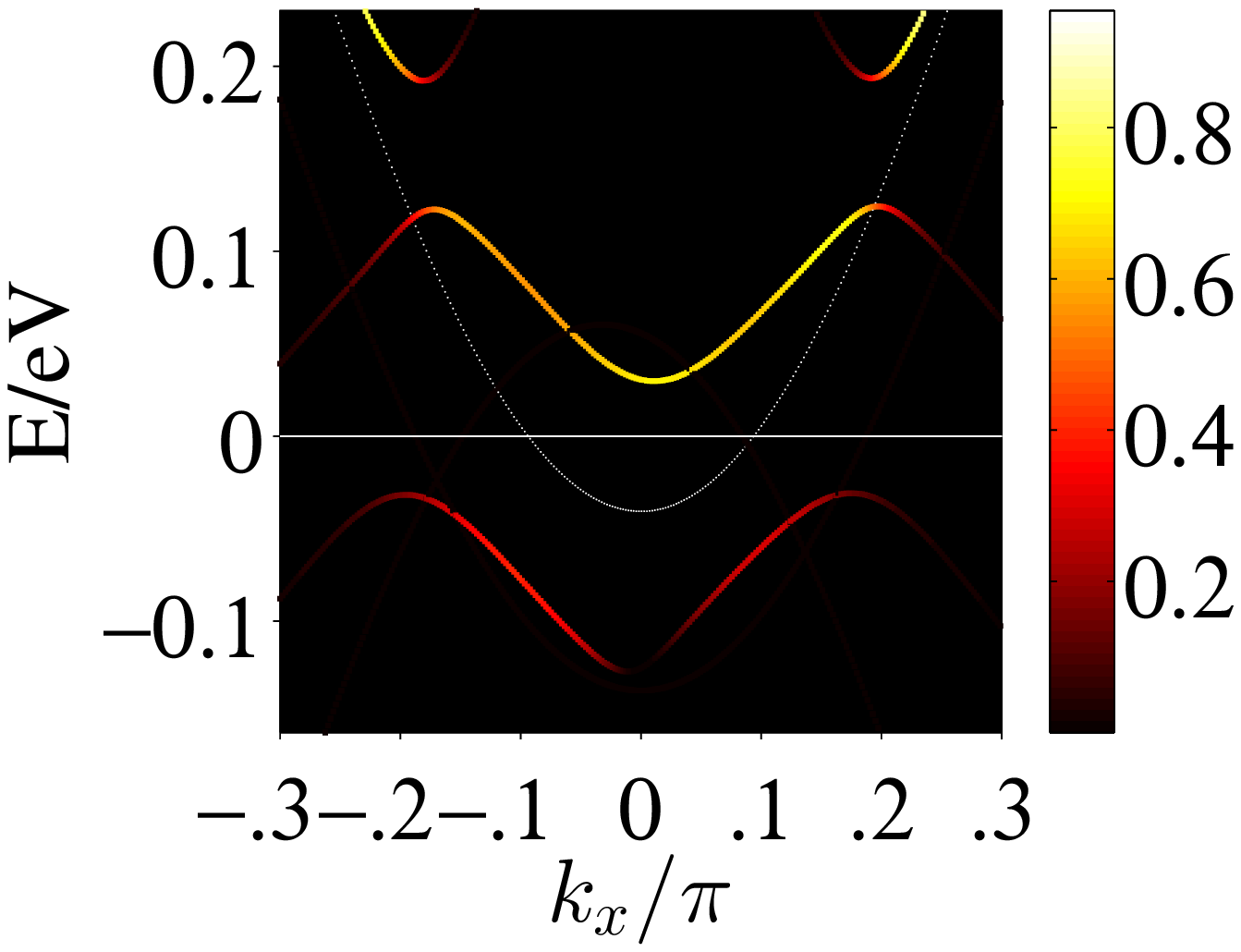}}
\caption{\label{fig: bands}(Color online) Quasi particle spectrum for the ME PDW state with $\delta K_y=0.1$. Shown are the bare electron dispersion
(the white parabola) and the PDW bands weighted by $|u(k)|^2$ (the negative energy portion is observable by ARPES). (a)  $\Delta_1=\Delta_2=75$ meV and $k_y=\pi$.
(b)  $\Delta_1=\Delta_2=75$ meV and $k_y=\pi-0.7$, here occupied bands have moved up to $\epsilon_F$ to create the Fermi arcs.
(c) $\Delta_1=85$ meV, $\Delta_2=65$ meV, and $k_y=\pi$. Notice the asymmetry in $k_x$ about $k_x=0$.}
\vspace{-5mm}
\end{figure}
\twocolumngrid

\noindent reversal symmetric PDW states are consistent with ARPES measurements. Here we focus our analysis on the tetragonal ME PDW state discussed in Section VI 
(qualitatively similar results will appear for the PDW state discussed in  Section VII). To examine the qp properties, we consider the Hamiltonian
\begin{equation}
H=\sum_{{\bm k},s} \epsilon_{{\bm k}}c^{\dagger}_{{\bm k}s} c_{{\bm k}s}+ \sum_{{\bm Q}_i,{\bm k}}[\Delta_{{\bm Q}_i}({\bm k})c^{\dagger}_{{\bm k}+\frac{{\bm Q}_i}{2},\uparrow} c^{\dagger}_{-{\bm k}+\frac{{\bm Q}_i}{2},\downarrow}\\+h.c.], \label{H}
\end{equation}
\noindent where $c_{{\bm k}s}$ is the fermion destruction operator with momentum ${\bm k}$ and spin $s$, $\epsilon_{\bm k}$ is the bare dispersion, and  $h.c.$ means 
Hermitian conjugate. We compute the eigenstates of Eq.~(\ref{H}) and the spectral weight using
\begin{figure}[t]
\begin{center}
\includegraphics[width=3.1in]{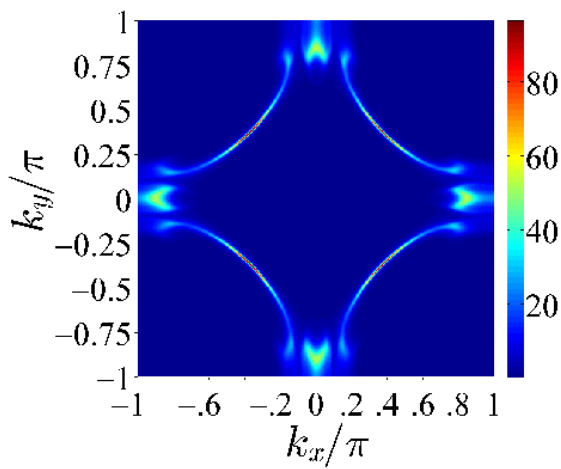}
\end{center}
\caption{(Color online)  Spectral weight showing Fermi arcs for ME PDW state. Here $\Delta_1=\Delta_2=75$ meV and $\Gamma=10$ meV.}
\label{fig3}
\vspace{-1mm}
\end{figure}
\vspace{-5mm}

\begin{equation}
I(\omega,{\bm k})= Im\sum_{\alpha} \frac{|u_{\alpha,{\bm k}}|^2}{w-E_{\alpha,{\bm k}}-i\Gamma},
\end{equation}
where $E_{\alpha,{\bm k}}$ are the eigenenergies of Eq.~(\ref{H}), $u_{\alpha,{\bm k}}$ is the weight of the fermion with momentum ${\bm k}$ in the band $\alpha$,
and the damping factor $\Gamma$ models short-range order in the PDW phase. In our calculations we use the bare dispersion $\epsilon_{\bm k}$ given in
Ref.~\onlinecite{he11} and set $\Gamma=0.1$ eV. In addition, we set $\Delta_{{\bm Q}_i}({\bm k})=\Delta_i f_i({\bm k}-{\bm K}_i)$ which localizes the pairing in
${\bm k}$ space as described in \cite{lee14} (for $\dq[1]$, $f_1({\bm k}-{\bm K}_1)=e^{-(k_y-K_y)^2/k_0^2}$, the other $f_i$ are determined by tetragonal symmetry).
Figures~\ref{fig: bands}\subref{subfig: banda} and \ref{fig: bands}\subref{subfig: bandb} show the bands weighted by a factor $|u_{\alpha,{\bm k}}|^2$ for fixed $k_y=\pi$ 
and $k_y=\pi-0.7$ as a function of $k_x$ (with $\Delta_1=\Delta_2$). These first two figures show that the Fermi arc results from occupied states moving towards the Fermi 
level, a point emphasized in Ref.~\onlinecite{lee14}. In Fig.~\ref{fig: bands}\subref{subfig: bandc} we illustrate the role of $\Delta_1\ne\Delta_2$. Notice that the ARPES 
bands become asymmetric about $k_x=0$. This asymmetry is consistent with existing ARPES measurements and it would be of interest to examine this experimentally. We note 
that this asymmetry does not exist in the PDW phase proposed in Ref.~\onlinecite{lee14}. Fig.~\ref{fig3} shows the spectral weight for $\Delta_1=\Delta_2=75$ meV revealing 
the Fermi arcs.

\vspace{-4mm}
\section{\label{sec: conclude} Conclusions}
\vspace{-2mm}
We have shown that PDW order can generate translational invariant ME loop current order as a secondary order parameter. We further show that there exists a PDW ground 
state with ME loop current order, CDW correlations, and qp properties consistent with ARPES. When phase fluctuations are included, a state appears in which only the ME
loop current order has long-range spatial correlations. We predict that this state will exhibit short-range incommensurate angular momentum correlations at the same
wavevector as the CDW correlations. We also show that this state gives rise to an asymmetry in the qp properties that may be observed by ARPES.
\vspace{-4mm}
\begin{acknowledgments}
\vspace{-2mm}
We thank Egor Babaev, Andrey Chubukov, Julien Garaud, Marc-Henri Julien, Patrick Lee, and Yuxuan Wang for fruitful discussions. We acknowledge support from NSF grant 
No. DMR-1335215.
\end{acknowledgments}

\end{document}